\documentclass[aps,prl,superscriptaddress,floatfix,twocolumn,showpacs]{revtex4-1}

\usepackage{verbatim}

\usepackage{amsmath}
\usepackage{amsfonts}
\usepackage{amssymb}
\usepackage{graphicx}
\usepackage{bm}
\usepackage{color}
\usepackage{epsf}

\usepackage{psfrag}

\def\bea{\begin{eqnarray}}
\def\eea{\end{eqnarray}}
\def\beas{\begin{eqnarray*}}
\def\eeas{\end{eqnarray*}}
\def\beqas{\begin{eqnarray*}}
\def\eqas{\end{eqnarray*}}
\def\beq{\begin{equation}}
\def\eeq{\end{equation}}
\def\beqd{\begin{displaymath}}
\def\eeqd{\end{displaymath}}
\def\eqd{\end{displaymath}}

\def\slashchar#1{\setbox0=\hbox{$#1$}
   \dimen0=\wd0
   \setbox1=\hbox{/} \dimen1=\wd1
   \ifdim\dimen0>\dimen1
      \rlap{\hbox to \dimen0{\hfil/\hfil}}
      #1
   \else\begin{eqnarray}
      \rlap{\hbox to \dimen1{\hfil$#1$\hfil}}
      /
   \fi}

\begin{document}
\title
{Diffractive deeply virtual Compton scattering}
\author{B.~Pire}
\affiliation{ CPHT, CNRS, \'Ecole Polytechnique, I.~P.~Paris,
 91128 Palaiseau,     France }
\author{L.~Szymanowski}
\affiliation{ National Centre for Nuclear Research (NCBJ), Pasteura 7, 02-093 Warsaw, Poland}
\author{S.~Wallon}
\affiliation{LPT, CNRS, Univ. Paris-Sud, Universit\'e Paris-Saclay, 91405, Orsay, France {\em \&} \\
Sorbonne Universit\'e, Facult\'e de Physique, 4 place Jussieu, 75252 Paris Cedex 05, France}

%\date{\today}

\begin{abstract}

\noindent
Diffractive deeply virtual Compton scattering (DiDVCS) is the process  $\gamma^*(- Q^2) + N \rightarrow \rho^0 + \gamma^* (Q'^2)+  N'$, where N is a nucleon or light nucleus,  in the kinematical regime of large rapidity gap between the $\rho^0$ and the final photon-nucleus system, and in the generalized Bjorken regime where both photon virtualities $Q^2$ and $ Q'^2$ are large. We show that this process has the unique virtue of combining the large diffractive cross sections at high energy with the tomographic ability of deeply virtual Compton scattering to scrutinize the quark and gluon content of nucleons and light nuclei. Its study at an electron-ion collider would enlighten the internal structure of hadrons.
\end{abstract}

\maketitle

\section{Introduction}

It is now common wisdom that the dominant mechanism of a diffractive electroproduction process in the hard regime is the scattering of a small transverse-size  ($O(\frac{1}{Q})$) colorless dipole on a nuclear target, where $Q$ is the virtuality of the exchanged hard photon. This justifies the use of perturbative QCD methods for the description of the process. In the  Regge inspired $k_T$ -factorization approach which is known to be applicable at high energy, $W \gg Q \gg \Lambda_{QCD}$, one writes the scattering amplitude in terms of two impact factors  with, at leading order, a two Reggeized gluon exchange in the t-channel. The Balitsky-Fadin-Kuraev-Lipatov (BFKL) evolution~\cite{Fadin:1975cb, Kuraev:1977fs, Balitsky:1978ic} can then be applied to account for specific large energy QCD resummation effects. 

On the other hand, the unique features of nearly forward exclusive hard scattering amplitudes in the generalized Bjorken regime allowed to construct a vast program aiming at the tomography of nucleons and light nuclei. The theoretical framework is collinear factorization~\cite{Mueller:1998fv,Ji:1998xh,Collins:1998be} of the scattering amplitude into generalized parton distributions~\cite{Diehl:2003ny, Belitsky:2005qn} and hard perturbatively calculable coefficient functions. Deeply virtual Compton scattering ($\gamma^*N \to \gamma N'$) (DVCS) and the timelike Compton scattering (TCS) related process ($\gamma N \to \gamma^* N'$) have been much discussed both theoretically and experimentally~\footnote{For recent reviews, see Refs.~\cite{Kumericki:2016ehc} and \cite{Anikin:2017fwu}
 and references therein.}, and shown to provide the best tool available for a 3-dimensional imaging~\cite{Burkardt:2000za, Ralston:2001xs, Diehl:2002he, Dupre:2016mai} of the quark and gluon structure of the proton and light nuclei.

The process we study here - called DiDVCS for diffractive deeply virtual Compton scattering - adds the merits of these two classes of reactions, with a large cross section at large energy and an excellent resolution of the nucleon's interior. It is particularly well suited for  future experiments at an electron-ion collider which is under active study recently~\cite{Boer:2011fh, AbelleiraFernandez:2012cc, Aschenauer:2013hhw}. 
\section{Kinematics}
%%%%%%%%%%%%%%%%%%%%%%%%%%%%%%%%%%
\label{Sec:Kinematics}

We study the process 
\begin{equation}
\gamma^*(q, \varepsilon) + N(p_1,\lambda_1) \rightarrow \rho^0(q_\rho,\varepsilon_\rho) + \gamma^*(q', \varepsilon') +  N'(p_2,\lambda_2)\,,
\label{processGeneral}
\end{equation}
at large squared energy $s_{\gamma N} =(q+p_1)^2$, in the forward limit where the $\rho$ meson flies in the same direction as the virtual initial photon and in the kinematical regime of large rapidity gap between the $\rho^0$ and the photon, {\em i.e.} $s_1=(q_\rho+q')^2 \gg s_2= (q'+p_2)^2$. 

  We define 
\begin{equation}
P^\mu = \frac{p_1^\mu + p_2^\mu}{2} ~,~ \Delta^\mu = p_2^\mu - p_1^\mu\,,
\end{equation}
and decompose momenta on a Sudakov basis  as
\begin{equation}
\label{sudakov1}
v^\mu = \gamma  n^\mu + \delta p^\mu + v_\bot^\mu \,,
\end{equation}
with $p$ and $n$ the light-cone vectors ($2p.n=s$)
\begin{equation}
\label{sudakov2}
p^\mu = \frac{\sqrt{s}}{2}(1,0,0,1)\,;  n^\mu = \frac{\sqrt{s}}{2}(1,0,0,-1) \,,
\end{equation}
and
\begin{equation}
\label{sudakov3}
v_\bot^\mu = (0,v^x,v^y,0) \,, \qquad v_\bot^2 = -\vec{v}_t^2\,.
\end{equation}
The particle momenta read
\begin{eqnarray}
\label{impini}
 q^\mu &=& n^\mu - \frac{Q^2}{s} p^\mu ~, \nonumber \\
 p_1^\mu &=& (1+\xi)\,p^\mu + \frac{M^2}{s(1+\xi)}\,n^\mu~, \nonumber\\
 p_2^\mu &=& (1-\xi)\,p^\mu + \frac{M^2+\vec{\Delta}^2_t}{s(1-\xi)}n^\mu + \Delta^\mu_\bot ~, \\
 p_\rho^\mu &=& \alpha_\rho \, n^\mu + \frac{m^2_\rho}{\alpha_\rho s}\,p^\mu \,,\nonumber\\
q'^\mu &=& \alpha \, n^\mu + \frac{Q'^2+\vec\Delta_t^2}{\alpha s}\,p^\mu  -\Delta^\mu_\bot~,\nonumber
\end{eqnarray}
with  
$M$ and $m_\rho$ the masses of the nucleon and of the $\rho$ meson.

The total squared center-of-mass energy of the $\gamma^*$-N system is
\begin{equation}
\label{energysquared}
S_{\gamma N} = (q + p_1)^2 \approx (1+\xi)s - Q^2 + M^2\,.
\end{equation}
Neglecting masses and $\Delta_T$, we have
\begin{eqnarray}
\alpha + \alpha_\rho = 1\,.
\end{eqnarray}
Moreover, in the large rapidity gap regime, we have
\begin{equation}
\alpha_\rho \approx  1 ; \alpha \ll  1 \,.
\end{equation}
The squared sub-energies are
\begin{eqnarray}
\label{forwards2}
s_1&=&(p_\rho +q')^2 \approx \frac{Q'^2}{\alpha} \,, \nonumber \\
s_2 &=& (p_2+q')^2 \approx s(1-\xi) \alpha +Q'^2\\
& \approx& S_{\gamma N} \frac{1-\xi}{1+\xi}\alpha + Q'^2\,.\nonumber
\end{eqnarray}
The skewness variable $\xi$ is thus expressed in terms of $Q'^2$ and of the squared sub-energies (we neglect here $Q^2$ compared to $s_{\gamma N} $) as :
\begin{eqnarray}
\label{forwardsxi}
\xi \approx \frac{Q'^2}{2s_2-Q'^2}\,.%\frac{1-\tau}{1+\tau}  ~~,~~\tau \approx \frac{s_1}{s_{\gamma N} } \frac{s_2-Q'^2}{Q'^2} \,.
\end{eqnarray}
%Note that contrarily to the DVCS or TCS cases, $\xi$ is not proportional to $Q'^2$. $\xi =1$ at threshold ($s_2=Q'^2$), and decreases  when the ratio $\frac{s_2}{Q'^2}$ grows. $\xi$ vanishes at the limiting value 
%$s_2 = Q'^2 [1+\frac{S_{\gamma N}}{s_1}]$.
%Remember that $0<\xi <1$ , {\i.e.} $0<\tau <1$ which means that $s_2 < Q'^2 [1+\frac{S_{\gamma N}}{s_1}]$.
 This quantifies our statement that this process probes the $\xi =O(1)$ region of GPDs when $ \frac{Q'^2}{s_2}= O(1)$.

\section{The scattering amplitude}

We write the scattering amplitude of the process 
\begin{equation}
\gamma_L^*(q,\varepsilon) + p(p_1,\lambda_1) \rightarrow  \rho^0_L(p_\rho) + \gamma_L^*(q', \varepsilon') + p(p_2,\lambda_2)\,
\label{process}
\end{equation}
 in the factorized form :

\begin{eqnarray}
\label{AmplitudeFactorized}
\mathcal{M} =\frac{i s}{N_c^2-1}  \int d^2k   \frac{ \Phi_1(k, r-k) \Phi_2(k, r-k)}{(2\pi)^2 k^2(r-k)^2},
\end{eqnarray}
where $\Phi_1(k, r-k)$ and $ \Phi_2(k, r-k)$ are the impact factors defined as  (the color sum being taken care of):
\begin{eqnarray}
\label{ImpactFact}
\Phi_1(k, r-k) &=&\frac{1}{2}\int \frac{d\delta}{2\pi} {\cal S}_{\mu \nu}^{\gamma^* g \to \rho g} \frac{2p^\mu p^\nu}{s} , \\ 
\Phi_2(k, r-k) &=&\frac{1}{2}\int \frac{d\gamma}{2\pi} {\cal S}_{\mu \nu}^{N g \to \gamma^* N' g} \frac{2n^\mu n^\nu}{s} , 
\end{eqnarray}
%%%%%%%%%%%%%%%%%%%%%%%%%%%%%%%%%%%%%%%%%%%%%%%%%%%%%
\begin{figure}
%[tb]
\center
\includegraphics[width=0.3\textwidth]{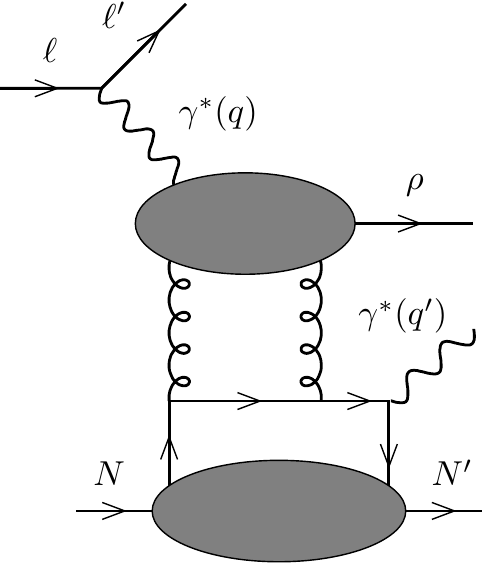}
 %\vspace{-1cm}
\caption{Diffractive  deeply virtual Compton scattering}
   \label{Fig1}
\end{figure}

\begin{figure}
\center
\includegraphics[width=0.4\textwidth]{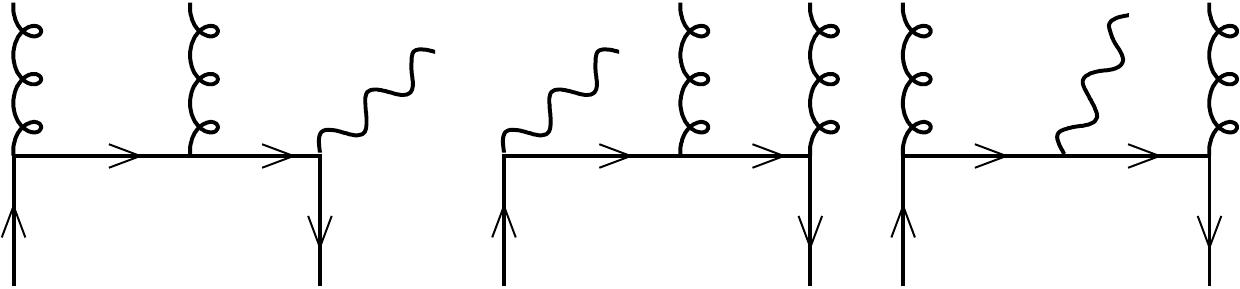}
\caption{Feynman diagrams for the impact factor $\Phi_2$ }
   \label{Fig2}
\end{figure}
When calculating the impact factors,  the gluon propagator numerators are replaced by $-g^{\mu\nu} \to \frac{-2 p^\mu n^\nu}{s}$ where $\mu$ (resp. $\nu$) is the  index acting on the upper (resp. lower) part of the Feynman diagram. 
The gauge invariance of the amplitude allows us to make the substitution $\varepsilon_L^\mu  \to \frac{2Q}{s} p^\mu$,  $\varepsilon'{}_L^{\mu}  \to \frac{2 Q'}{\alpha s} p^\mu$ (in the forward limit $\Delta_\perp=0$).

The $\gamma^*_L \to \rho_L$  impact factor is well known~\cite{Ginzburg:1985tp, Ivanov:2004pp} and reads at leading order (for $\vec r =0$) :
\begin{eqnarray}
\label{ImpactFacts}
 \Phi_1(\vec k, -\vec k) = \frac{C_1}{Q} \int  \frac{du  \phi(u) \vec k^2}{\vec k^2+u (1-u) Q^2}\, ,
\end{eqnarray}
where  $C_1=2C_F  \sqrt{4\pi\alpha_{em}} (4\pi \alpha_s) \frac{ f_{\rho^0}}{\sqrt{2}}$, $e_u=2/3$, $e_d=-1/3$ and $\phi(u)$ is the longitudinally polarized $\rho^0$ meson distribution amplitude 
defined, at the leading twist 2, by the matrix element~\cite{Ball:1996tb}
\begin{equation}
\langle 0|\bar{u}(0)\gamma^\mu u(x)|\rho^0(p_\rho,\varepsilon_{\rho L}) \rangle = \frac{1}{\sqrt{2}}p_\rho^\mu f_{\rho^0}\!\!\!\int_0^1 \!\! du\ e^{-iup_\rho\cdot x} \phi(z),
\label{defDArhoL}
\end{equation}
with $f_{\rho^0}=216\,\mbox{MeV}$, and by a similar expression with opposite sign for d quarks. 

The lowest impact factor $\Phi_2$ is a new object that we calculate in the collinear approximation from the diagrams of Fig.~\ref{Fig2}. It vanishes at leading twist for a transverse virtual photon and equals for a longitudinally polarized virtual photon  to:
\begin{eqnarray}
\label{IF2}
&&\Phi_2(\vec k, -\vec k) = \sum _q\frac{C_2 }{Q'}  \int_{-1}^1 dx ~\theta(\xi^2-x^2)~ 4\xi^2 \vec k^2  \\
&&\times \frac{ \bar U(p_2,\lambda_2) [ \hat n e_qH^q(x,\xi,t) +\frac{i \sigma^{n\Delta}}{2M}e_qE^q(x,\xi,t) ]U(p_1,\lambda_1)}{ (x^2-\xi^2) Q'^2 +4 \xi^2 \vec k^2 -i\epsilon } \,, \nonumber
\end{eqnarray}
with  $C_2= - 4C_F (4\pi \alpha_s )\sqrt{4\pi\alpha_{em}}/s$ and where $U(p_i,\lambda_i)$ are the in-going and out-coming nucleon spinors 	and $H(x,\xi,t)$ and $E(x,\xi,t)$ are the usual vector generalized parton distributions defined as~\cite{Mueller:1998fv}:
\begin{eqnarray}
&&\langle p(p_2,\lambda')|\, \bar{q}\left(-\frac{y}{2}\right)\,\gamma^+q \left(\frac{y}{2}\right)|p(p_1,\lambda) \rangle  \nonumber \\
&&= \int_{-1}^1dx\ e^{-\frac{i}{2}x(p_1^++p_2^+)y^-}\bar{u}(p_2,\lambda')\, \left[ \gamma^+ H^{q}(x,\xi,t)  \right.\nonumber \\
&& ~~~~~~~~ +\left. \frac{i}{2m}\sigma^{+ \,\alpha}\Delta_\alpha  \,E^{q}(x,\xi,t) \right]
u(p_1,\lambda)\, .
\label{defGPD}
\end{eqnarray}
Note that there is no contribution to the impact factor from the axial nor from the transversity GPDs and that the  contributions  of the $H$ and $E$ GPDs to the impact factor come only from their ERBL region $-\xi < x < \xi$. This property is rather unique and much related to the diffractive kinematics as shown in the related studies~\cite{Ivanov:2002jj, Enberg:2006he,Cosyn:2019eeg}.There is of course no contribution from the gluon GPDs since the C-parity of the (two gluon + photon) system is odd.

This impact factor looks very much like $\Phi_1$ after a suitable redefinition of the integration variable as $u=\frac{x+\xi}{2\xi}$, which is not surprising since the lower impact factor is very similar to the upper one when the $\rho$ meson DA is replaced by the GPD. It has however a different and  very interesting analytic structure; it develops an imaginary part when $\vec k^2 < \frac{Q'^2}{4} $. This imaginary part  is proportional to the value of the GPDs at 
$x= \pm \xi \sqrt {1-\frac {4\vec k^2} {Q'^2}}$. This is reminiscent of, but different from, the imaginary part developed by the DVCS amplitude for $x= \pm \xi$. The difference is due to the non-vanishing virtuality of the $t-$channel exchanged gluons, in the present $k_T-$factorization approach, as compared with the usual collinear factorization.

\begin{figure}
%[tb]
\center
\includegraphics[width=0.4\textwidth]{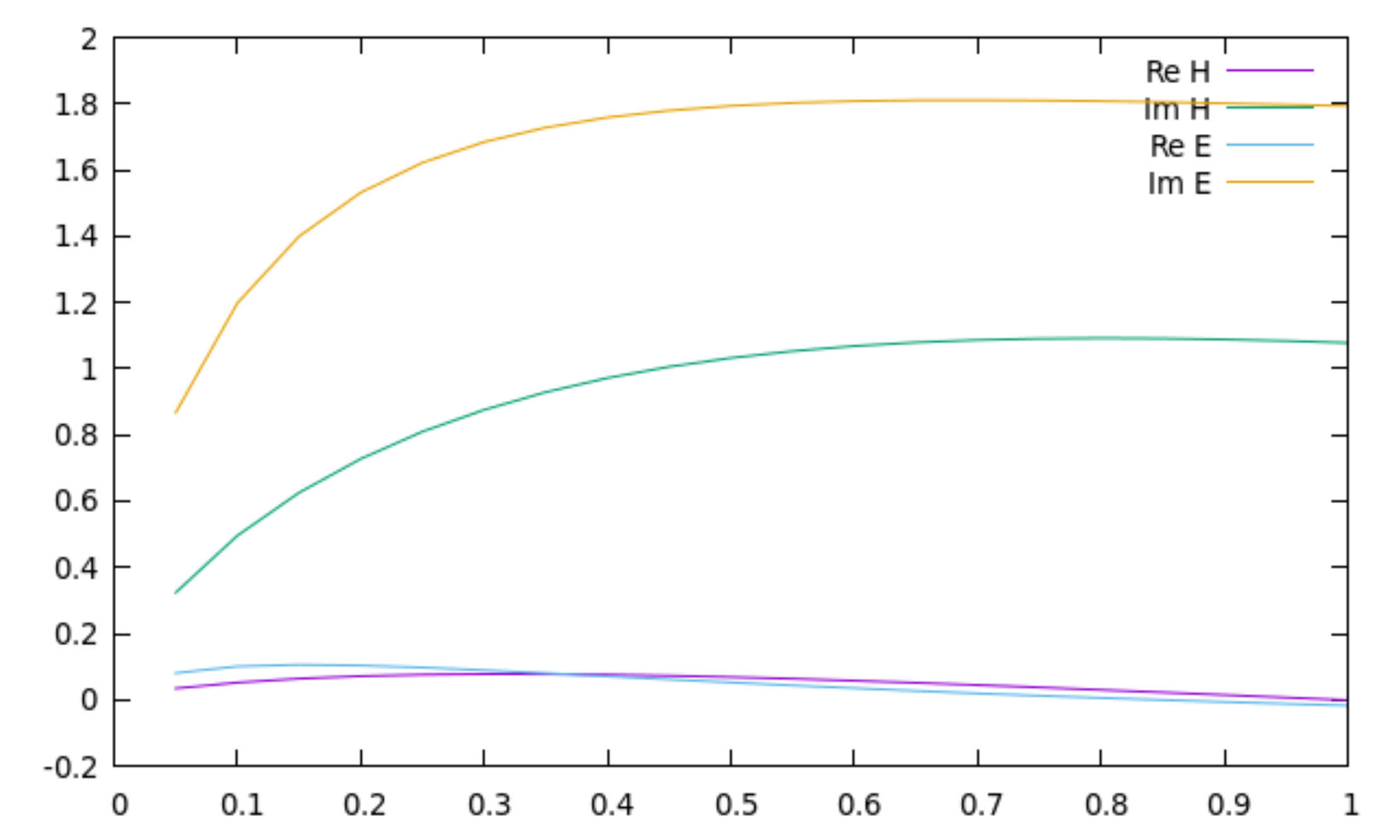} 
\caption{An estimate of the real (lower curves) and  imaginary (upper curves) parts of the diffractive Compton form factors ${\cal H}^d$ (left panel) and  ${\cal E}^d$ (right panel) at $Q^2=Q'^2= 4~{\rm GeV}^2$ and $t = -0.1$GeV$^2$, divided by the pre-factor $6f_\rho K/\sqrt 2 Q^4$ from Eq.~(\ref{CFFint}).}
   \label{Fig3}
\end{figure}

The $\gamma^*_L N \to \rho_L \gamma^*_L N'$ amplitude in the diffractive region can then be written in terms of {\em diffractive Compton form factors} ${\cal H}^d(\xi,t)$ and ${\cal E}^d(\xi,t)$ as
\begin{eqnarray}
\label{Amplitude}
\hspace{-.4cm} i\mathcal{M} = \bar U(p_2,\lambda_2) [ \hat n {\cal H}^d(\xi,t) +\frac{i \sigma^{n\Delta}}{2M} {\cal E}^d(\xi,t) ]U(p_1,\lambda_1),\,\,
\end{eqnarray}
where 
\begin{eqnarray}
\label{CFF}
{\cal H}^d(\xi,t) &=& C \! \int \! \frac{d^2\vec k}{(2\pi)^2(\vec k^2)^2} \int_0^1 du  \frac{\vec k^2 \phi(u)}{\vec k^2+Q^2 u (1-u)} \, \, \\
&&\times \int_{-\xi}^\xi dx  \frac{\vec k^2 [e_uH^u(x,\xi,t)+ e_d H^d(x,\xi,t)]}{\vec k^2+Q'^2 \frac{x^2-\xi^2}{4\xi^2} -i\epsilon}  \,, \nonumber \\
 {\cal E}^d(\xi,t) ]&=& C \! \int \! \frac{d^2\vec k}{(2\pi)^2(\vec k^2)^2} \int_0^1 du  \frac{\vec k^2 \phi(u)}{\vec k^2+Q^2 u (1-u)} \, \\
 && \times \int_{-\xi}^\xi dx  \frac{\vec k^2 [e_uE^u(x,\xi,t)+ e_dE^d(x,\xi,t)]}{\vec k^2+Q'^2 \frac{x^2-\xi^2}{4\xi^2} -i\epsilon} \,, \nonumber
\end{eqnarray}
with $C = \frac{2^9 \pi^3 \alpha_{em} C_F^2 \alpha_S^2 f_{\rho}}{\sqrt{2} Q Q' (N_c^2-1)}.$
In the simple case where the $\rho$ DA has the asymptotic shape $\Phi(u)= 6 u (1-u)$, the $d^2\vec k$ and the $du$ integrations can be performed analytically, while the  $dx$ integration depends on the specific shape of the nucleon GPDs~\footnote{We have checked numerically that the impact factor $\Phi_1(\vec k^2)$ does not depend much on the shape of the $\rho$ DA. Replacing the asymptotic shape by a $\sqrt{u(1-u)}$~\cite{Forshaw:2012im} only modifies by a few per cent its magnitude, and only in the small $\vec k^2$ region.}.
After the integration over $d^2\vec k$, the diffractive Compton form factor ${\cal H}^d$   reads in this case:
\begin{eqnarray}
\label{CFFint}
&&{\cal H}^d = \frac{6 f_\rho K}{\sqrt{2}Q^3Q'}\!\int_{-\xi}^\xi \! dx  [ e_u H^u(x,\xi,t) + e_d H^d(x,\xi,t)  ]  \\
&&\hspace{0mm} \times \left\{ \frac{-2z}{\sqrt{4z+1}} \left[{\rm Li}_2\left(\!\frac{-2}{\sqrt{4z+1}-1}\right)- {\rm Li}_2\left(\!\frac{2}{\sqrt{4z+1}+1}\right)\right]  \right.\!\!\!\nonumber \\
&&\hspace{0cm}\left.-2-( \ln z -i\pi)\left[ 1-\frac{4z}{\sqrt{4z+1}} \coth^{-1} (\sqrt{4z+1}) \right]  \right\}, \nonumber
\end{eqnarray}
with $z= \frac{Q'^2}{Q^2} \frac{ (\xi^2-x^2)}{4 \xi^2}$ and $K= \frac{2^7 \pi^2 \alpha_{em}C_F^2\alpha_S^2}{N_C^2-1}$, and a similar expression for ${\cal E}$.
We show on Fig.~\ref{Fig3} the $\xi -$dependence of the real and imaginary parts of these diffractive form factors in the case $Q^2=Q'^2$, with the GPDs taken from Ref.~\cite{Goloskokov:2006hr}.

\section{Next to leading order corrections}
The amplitude (\ref{Amplitude}) is valid at the leading order in $\alpha_s$ and one must discuss how the inclusion of higher order contributions will modify it. We leave for further work these delicate studies but 
summarize what is known about these contributions in quite similar processes, and what we might expect from their study.
\begin{itemize}
\item{Firstly, the $\gamma^* \rho$ impact factor is known at NLO~\cite{Ivanov:2004pp} in the forward kinematics discussed here~\footnote{This has been recently extended in arbitrary kinematics~\cite{Boussarie:2016bkq}.} and there is no basic problem to include the NLO result here. The inclusion of this
result in the  process  $\gamma^*\gamma^*\to \rho \rho$ lead to important numerical corrections~\cite{Ivanov:2005gn}.}
\item{The two gluon exchange is but the first step in the construction of the perturbative Pomeron and BFKL evolution should be implemented to take care of the tower of reggeized gluon exchanges. This has been performed for the simpler  process  $\gamma^*\gamma^*\to \rho \rho$~\cite{Enberg:2005eq,Ivanov:2005gn}, leading to a big enhancement factor with a moderate energy dependence when compared to a two gluon exchange~\cite{Pire:2005ic,Segond:2007fj}.}
\item{Finally, the Born approximation calculation of the $N \to \gamma^* N'$ impact factor should be supplemented with one loop calculation. The similarity of the Born results for $\gamma^* \rho$ and $N \to \gamma^* N'$ impact factors naturally leads to the conjecture of a very similar NLO analysis for both cases, with the important difference of the {\em timelike} vs {\em spacelike} nature of the virtual photon, where one expects analytic continuation effects, such as those discussed in Ref.~\cite{Muller:2012yq}.}
\end{itemize}

\section{Phenomenological perspectives}
Eqs.~(\ref{Amplitude}-\ref{CFFint}) are the central results of our study. Let us now discuss the main signatures of our description of diffractive DVCS.
\begin{itemize}
\item{The amplitude is energy independent at Born order, and would acquire a mild energy dependence when BFKL evolution is turned on.}
\item{The amplitude scales like $\frac{S_{\gamma N}}{Q^3Q'} f(Q/Q',\xi)$, at fixed $t$.}
\item{The initial virtual photon is longitudinally polarized, which makes the electroproduction cross section $\varphi-$independent, where $\varphi$ is the azimuthal angle of the initial lepton plane. }
\item{The produced virtual photon is longitudinally polarized. The lepton pair that originates from its decay will thus have the characteristic $\sin^2 \theta$ shape, where $\theta$ is the angle of one lepton momentum with respect to the dilepton momentum, boosted to the dilepton center of mass system.}
\item{There is a leading target transverse spin asymmetry, proportional to the product of the ${\cal H}^d(\xi,t) $ and ${\cal E}^{d*}(\xi,t)$ diffractive Compton form factors.}
\end{itemize}
We leave for further studies a detailed phenomenological analysis of the diffractive DVCS reaction. (Note added : After completing a phenomenological study of the process  (see W.~Cosyn and B.~Pire,
[arXiv:2103.01411 [hep-ph]]), this process seems very difficult to  measure at future electron-ion colliders with their nominal luminosities).

%%%%%%%%%%%%%%%%%%%%%%%%%%%%%%%%%%%%%%%%%%%%%%%%%%%%%%%%%%%%%%%%%%%%%%%%%%%%%%%%%%%%%%%%%%% 
\paragraph*{Acknowledgements.} 

\noindent
We thank Renaud Boussarie, Wim Cosyn and St\'ephane Munier  for   useful discussions.  This project has received funding from the European Union's Horizon 2020 research and innovation programme under grant agreement No 824093. L. S. is supported by the grant 2017/26/M/ST2/01074 of the National Science Center in Poland. He thanks CNRS,  the LABEX P2IO, the  GDR QCD  and the French-Polish Collaboration Agreement POLONIUM for support. 

%%%%%%%%%%%%%%%%%%%%%%%%%%%%%%%%%%%%%%%%%%%%%%%%%%%%%%%%%%%%%%%%%%%%%%
  
%

\end{document}